Correspondence and requests for materials should be addressed to A.Z. (andrej.zorko@ijs.si)

# Strain-Induced Extrinsic High-Temperature Ferromagnetism in the Fe-Doped Hexagonal Barium Titanate

A. Zorko[1,*], M. Pregelj[1], M. Gomilšek[1], Z. Jagličić[2,3], D. Pajić[4], M. Telling[5,6], I. Arčon[1,7], I. Mikulska[7], and M. Valant[7]

[1]Jožef Stefan Institute, Jamova c. 39, SI-1000 Ljubljana, Slovenia, [2]Institute of Mathematics, Physics and Mechanics, Jadranska c. 19, 1000 Ljubljana, Slovenia, [3]Faculty of Civil and Geodetic Engineering, University of Ljubljana, Jamova c. 2, SI-1000 Ljubljana, Slovenia, [4]Department of Physics, Faculty of Science, University of Zagreb, Bijenička c. 32, HR-10000 Zagreb, Croatia, [5]ISIS Facility, Rutherford Appleton Laboratory, Chilton, Didcot, Oxon OX11 OQX, United Kingdom, [6]Department of Materials, University of Oxford, Parks Road, Oxon, UK, [7]University of Nova Gorica, Vipavska 13, SI-5000 Nova Gorica, Slovenia.

**Diluted magnetic semiconductors possessing intrinsic static magnetism at high temperatures represent a promising class of multifunctional materials with high application potential in spintronics and magneto-optics. In the hexagonal Fe-doped diluted magnetic oxide, 6H-BaTiO$_{3-\delta}$, room-temperature ferromagnetism has been previously reported. Ferromagnetism is broadly accepted as an intrinsic property of this material, despite its unusual dependence on doping concentration and processing conditions. However, the here reported combination of bulk magnetization and complementary in-depth local-probe electron spin resonance and muon spin relaxation measurements, challenges this conjecture. While a ferromagnetic transition occurs around 700 K, it does so only in additionally annealed samples and is accompanied by an extremely small average value of the ordered magnetic moment. Furthermore, several additional magnetic instabilities are detected at lower temperatures. These coincide with electronic instabilities of the Fe-doped 3C-BaTiO$_{3-\delta}$ pseudocubic polymorph. Moreover, the distribution of iron dopants with frozen magnetic moments is found to be non-uniform. Our results demonstrate that the intricate static magnetism of the hexagonal phase is not**

**intrinsic, but rather stems from sparse strain-induced pseudocubic regions. We point out the vital role of internal strain in establishing defect ferromagnetism in systems with competing structural phases.**

The search for dilute magnetic oxides (DMOs) is at the forefront of spintronics and magneto-optics research and application[1–6]. The great interest has been triggered by the possibility of combining diverse functionalities of semiconductor electronics and magnetism in a single material, which would boost its application potential. Existence of magnetoelectric coupling in wide-band-gap semiconducting oxides at room temperature would be of great technological importance, given that the magnetic ordering is indisputably intrinsic. Indeed, a number of studies have claimed such intrinsic ordering has been observed in non-magnetic oxide semiconductors lightly doped with paramagnetic transition-metal ions. These findings are, however, a subject of controversy[7–9]. Very often the corresponding studies are focused only on routine characterization and modelling of the material's properties rather than critical analysis of the origin of the observed static magnetism. Recently, more detailed experiments have proven wrong a significant part of the claims about intrinsic magnetism in DMOs[10–14]. Moreover, aggregation of magnetic ions that leads to chemical phase separation on a nano-scale, also known as spinodal decomposition into regions with high/low concentration of dopants and a crystal structure imposed by the semiconducting host, has been lately witnessed in various cases[2,3,15-17]. Thus, the need for careful analysis of the magnetic properties of these systems at the microscopic level and their critical assessment are of paramount importance.

Since the discovery[18] of room-temperature ferromagnetism in the bulk Fe-doped hexagonal[19] $6H$-BaTiO$_{3-\delta}$ (Figure 1) that can be stabilized from the pseudocubic (3$C$) perovskite structure (Figure 1) by transition-metal doping[20], this system has been in the focus of investigations. Although simultaneous magnetic and polar orders (multiferroicity), highly desired for device applications, cannot be achieved in this material (1% Fe doping destroys ferroelectric order[18]), the magnetoelectric coupling has recently been demonstrated[21]. Moreover, arguments speaking in favour of its intrinsic ferromagnetism would make Fe-doped 6$H$-

BaTiO$_{3-\delta}$ rather special in the field of ferromagnetic (FM) DMOs. In contrast to many other materials where dopant-ion segregation and formation of precipitates are regularly encountered, the very high solubility of 6H-BaTiO$_3$ for Fe (and other transition metals), the possibility of growing single-crystal DMOs, and the apparent absence of any spurious precipitates all seem as strong evidences for its intrinsic FM character[18,22,23].

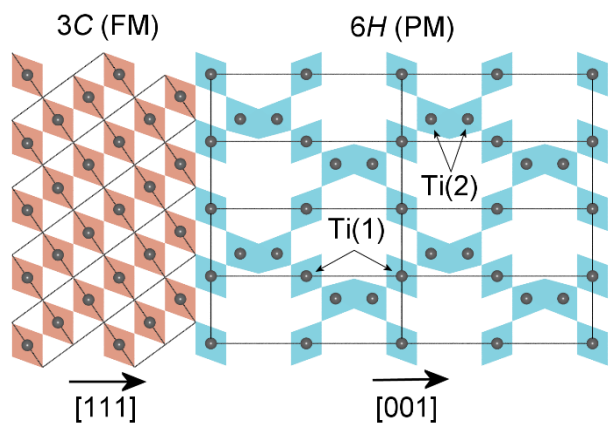

**Figure 1: Schematic interface of the two crystallographic phases in BaTiO$_{3-\delta}$.** Pseudocubic (3C) and hexagonal (6H) crystallographic polymorph of BaTiO$_{3-\delta}$. Fe$^{3+}$ cations substitute for Ti$^{4+}$ ions (spheres) within FeO$_6$ ochatedra (polyhedra). Two crystallographically different Ti sites are found in the 6H structure, while only one exists in the 3C structure. High-temperature ferromagnetism (FM) is ascribed to sparse regions of the 3C phase, while the majority 6H phase remains paramagnetic (PM).

The FM state of the Fe-doped 6H-BaTiO$_{3-\delta}$ is, however, highly unusual, if not even controversial. While the amount of the room-temperature FM signal significantly decreases with increasing dopant concentration[24], the average value of the static magnetic moment per Fe$^{3+}$ ion never exceeds a few hundredths of a Bohr magneton ($\mu_B$); compared to the full moment of 5$\mu_B$. Different hypotheses about such a drastic moment reduction have been put forward in the literature. These include competition between superexchange interactions leading to intrinsic magnetic inhomogeneity in the system[18] or, alternatively, the formation of dispersed Fe$^{3+}$ clusters around oxygen vacancies[24]. Furthermore, the size of the average FM moment substantially depends on various factors, including temperature, atmosphere[21,25], and duration of the synthesis[26]. Also the inevitable oxygen vacancies in the Fe-doped 6H-BaTiO$_{3-\delta}$ material, which compensate for the charge mismatch between $Fe^{3+}$ and $Ti^{4+}$, are suggested[21,23] to play an intrinsic role in establishing ferromagnetism through a dynamical exchange of trapped electrons among bound polarons[27]. However, the corresponding reports are contradictory, since the FM response can apparently be enhanced in either reduced samples[22] or in samples annealed in oxygen atmosphere[25]. Moreover, it has recently been suggested that the FM behaviour of the Fe-doped 6$H$-BaTiO$_{3-\delta}$ should rather be related to thermally activated diffusion, leading to a cation order-disorder transition, and that the segregation of oxygen vacancies plays no major role in ferromagnetism[28]. The resulting clustering of $Fe^{3+}$ ions at the two neighbouring pentahedral Ti(2) sites (Figure 1) and the segregation of oxygen vacancies at the connecting oxygen site[29], lead to the formation of lattice-matched Fe$_2$O$_8$ dimers. These have been predicted by ab-initio calculations[18], and are regularly observed in experiments[28,30,31], further revealing that a major portion of the Fe dopants are involved in their formation[28,31]. This fact thus evidently questions the cation ordering as the intrinsic origin of ferromagnetism in 6$H$-BaTiO$_{3-\delta}$, given the extremely small size of the average value of the magnetic moment per Fe dopant.

The FM behaviour of the Fe-doped 6$H$-BaTiO$_{3-\delta}$ has not yet been explored above room temperature or examined by local-probe magnetic spectroscopy techniques, which could potentially help in identifying any spurious effects. Therefore, we performed a comprehensive magnetic investigation, combining both bulk magnetic measurements as well as the complementary local-probe measurements of electron spin resonance (ESR) and muon spin relaxation ($\mu$SR), in the broad temperature range between 2 and 850 K. Here, we report a plethora of magnetic anomalies that appear in the Fe-doped 6$H$-BaTiO$_{3-\delta}$ below ~700 K. These coincide with the electronic instabilities of the Fe-doped pseudocubic 3$C$-BaTiO$_3$ polymorph. We provide firm experimental evidence that the previously reported static magnetism is not an intrinsic property of the doped 6$H$-BaTiO$_3$ hexagonal phase, but is instead related to sparse regions of strain-induced pseudocubic phase, which arise as a consequence of locally competing structural phases.

**Results**

In order to provide clear evidence for either intrinsic or extrinsic origin of magnetism in the Fe-doped 6$H$-BaTiO$_{3-\delta}$, an in-depth experimental study, which combines bulk and local-probe magnetic characterization techniques, is necessary. All our measurements were performed on high-quality polycrystalline samples (see Methods); some of which were additionally annealed. This annealing proves crucial for the existence of FM behaviour. In what follows, we label the non-annealed and annealed samples with F$c$BTO and F$c$BTOa, respectively. Here $c$ denotes the doping percentage of Fe$^{3+}$ ions at the Ti sites.

**Detection of magnetic instabilities by bulk magnetic measurements**

Bulk magnetization ($M$) measurements display a clear difference between the non-annealed and annealed samples (Figure 2a). While the latter exhibit complicated zero-field cooled (ZFC)/field-cooled (FC) split $M(T)$ curves, a paramagnetic (PM) phase is found in the former. The PM phase is characterized by a monotonic temperature dependence and absence of the ZFC/FC bifurcation. The corresponding molar susceptibility (per mole Fe) for the non-annealed samples, $\chi_{mol} = M/H$ ($H$ denotes the applied magnetic field), agrees well with the Curie-Weiss model $\chi_{mol} = N_A \mu^2 / 3 k_B (T - \theta_{CW})$. Here $N_A$ and $k_B$ denote the Avogadro and the Boltzman constant, respectively, $\theta_{CW}$ is the Weiss temperature and $\mu = 4.6(1)\mu_B$ is the average magnetic moment per Fe dopant. The derived magnetic moment is common to both F10BTO and F20BTO samples and is somewhat reduced from the value of $g\mu_B\sqrt{S(S+1)} = 5.92\mu_B$ expected for Fe$^{3+}$ ($S$ = 5/2) moments. However, the Fe$^{3+}$ valence was clearly revealed by X-ray absorption near-edge structure (XANES) spectroscopy[28,29] as being by far dominant in 6$H$-BaTiO$_{3-\delta}$. For F10BTO and F20BTO samples we find antiferromagnetic (AFM) Weiss temperature of -3(1) K and -7(2) K, respectively. This parameter scales linearly with the doping concentration and provides the energy scale of the average exchange coupling $J$ between the iron dopants. We find $J$ to be rather small compared to room temperature, where the FM state has been reported. Secondly, it is AFM and thus questions the intrinsic origin of the FM order.

At high temperatures ($T > T_{c1} \sim 700$ K) the $M(T)$ curves of the annealed samples match those of the non-annealed ones, proving the same PM state. This behaviour, however, drastically changes below $T_{c1}$, especially in the low applied field of 10 mT (Figure 2a). On decreasing temperature a ZFC/FC splitting is observed in F20BTOa at $T_{c1}$, with the FM component being enhanced below $T_{c2} \sim 570$ K, and the ZFC/FC difference increased below $T_{c3} \sim 450$ K. Finally, an additional magnetic anomaly, reflected in a local magnetization maximum, is found at $T_{c4} \sim 110$ K. The $M(H)$ magnetization curves of the same sample (Figure 2b) clearly display the same high-temperature magnetic transitions, as a hysteresis appears below $T_{c1}$ and is enhanced below $T_{c2}$. The deduced average value of the FM ordered magnetic moment ($\mu_{FM} = 1.5 \cdot 10^{-2} \mu_B$ per Fe at 300 K; see Methods) is dramatically reduced from the expected full $5\mu_B$ $Fe^{3+}$ value. Moreover, the $M(H)$ curves change linearly with $H$ at larger fields; a response being typical for PM and not FM systems. Therefore, below $T_{c1}$ one can think of the system as being predominantly PM with an additional small FM component. Such a simple picture is confirmed by $M(T)$ data in higher magnetic fields, where the high-temperature magnetic anomalies, although appearing at the same critical temperatures, become drastically suppressed (Figure 2a) due to saturation of the FM component. Similarly, as regularly reported before[18,23], the PM component that shows a Curie-Weiss temperature dependence dominates the magnetic response at low-temperatures (see the inset in Figure 2b), as the splitting of the FM hysteresis is only moderately enhanced at 2 K compared to room temperature (Figure 2b). The PM component is far from being saturated at low temperatures even at 5 T, similarly as found before[18,22,23]. The average value of the PM magnetic moment per $Fe^{3+}$ ion can be modelled[32] with a modified Brillouin function for spin 5/2, $M = f \cdot \mu \cdot B_{5/2}(\mu B / k_B T_{eff})$, by renormalizing the moment ($\mu = 5/2 \cdot g\mu_B$; $g$ denotes the Landé $g$-factor) with a factor $f$ and introducing an effective temperature $T_{eff} = T + T_0$. The fit to the FBTO20a data yields the empirical temperature $T_0 = 2.1$ K (inset in Figure 2b), confirming that AFM interactions are pertinent to this phase[32]. As the actual concentration of iron dopants in all our samples is very close to the nominal one (see Methods), the obtained renormalization factor $f = 0.25$ reveals that the majority of spins do not contribute to the magnetic signal at low temperatures. Such renormalization factor that has been found to decrease steadily with the doping concentration[18] can then be accounted for by formation of non-magnetic AFM entangled iron pairs at low temperatures, which are correlated with the formation of the $Fe_2O_8$ structural dimers[28,31]. The bulk-magnetization results thus imply the presence of FM and PM magnetic phases in the annealed 6H-BaTiO3-$\delta$ samples below $T_{c1}$ and corroborate impurity-based ferromagnetism, as further verified below by local-probe magnetic characterizations.

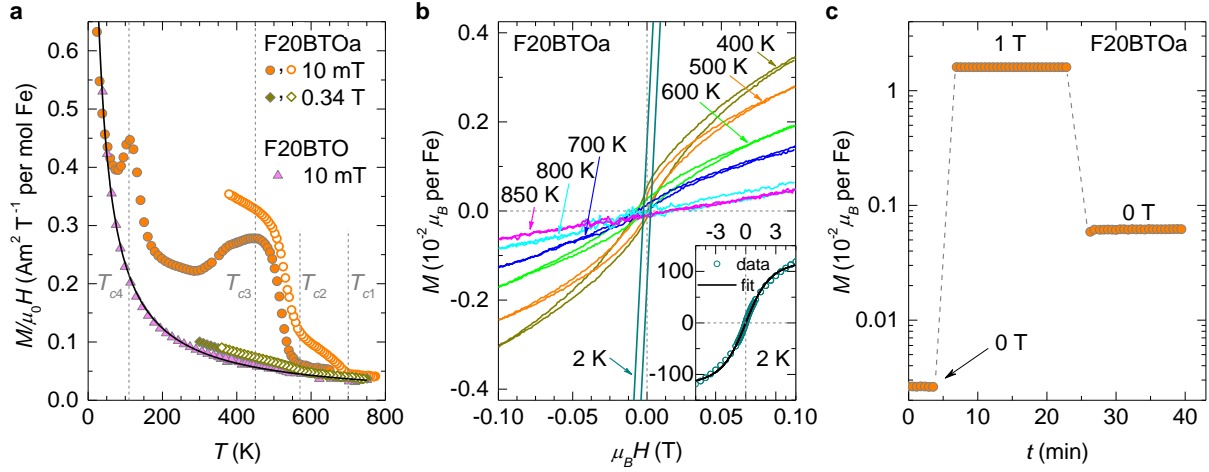

**Figure 2: Bulk magnetization measurements.** (**a**) The temperature dependence of zero-field cooled (full symbols) and field-cooled (open symbols) field-normalized bulk magnetization in the 20% Fe-doped 6$H$-BaTiO$_{3-\delta}$. The solid line is the fit to the Curie-Weiss model $M/\mu_0 H = N_A \mu^2 / 3k_B (T - \theta_{CW})$ (see text for details), while the vertical dashed lines display the temperatures of the magnetic transitions in the annealed sample. (**b**) Magnetization curves of F20BTOa at different temperatures displaying the FM hysteresis. Inset shows the 2 K data (circles) and the fit with the modified Brillouin function $M = f \cdot \mu \cdot B_{5/2}(\mu B / k_B T_{eff})$ (see text for details). (**c**) The time-dependent magnetization in the F20BTOa sample at 300 K after applying and removing the magnetic field of 1 T.

**Peculiarities of FM resonance modes**

Further insight to the unusual magnetism of the Fe-doped BaTiO$_{3-\delta}$ samples on a local scale is provided by ESR. Again, drastically different ESR spectra are found for the non-annealed and annealed samples. The former are featureless single-line ESR spectra, while the latter possess several lines that significantly evolve with temperature below $T_{c3}$ (Figure 3a). We note that the ESR spectra of the PM non-annealed samples are different from the fine-structured ESR spectra reported in previous studies[22,33–35]. However, it should be stressed that all previous studies were limited to doping concentrations below 5%. The disappearance of the fine structure in our non-annealed samples suggests enhanced magnetic interactions (dipolar and/or exchange) between iron dopants.

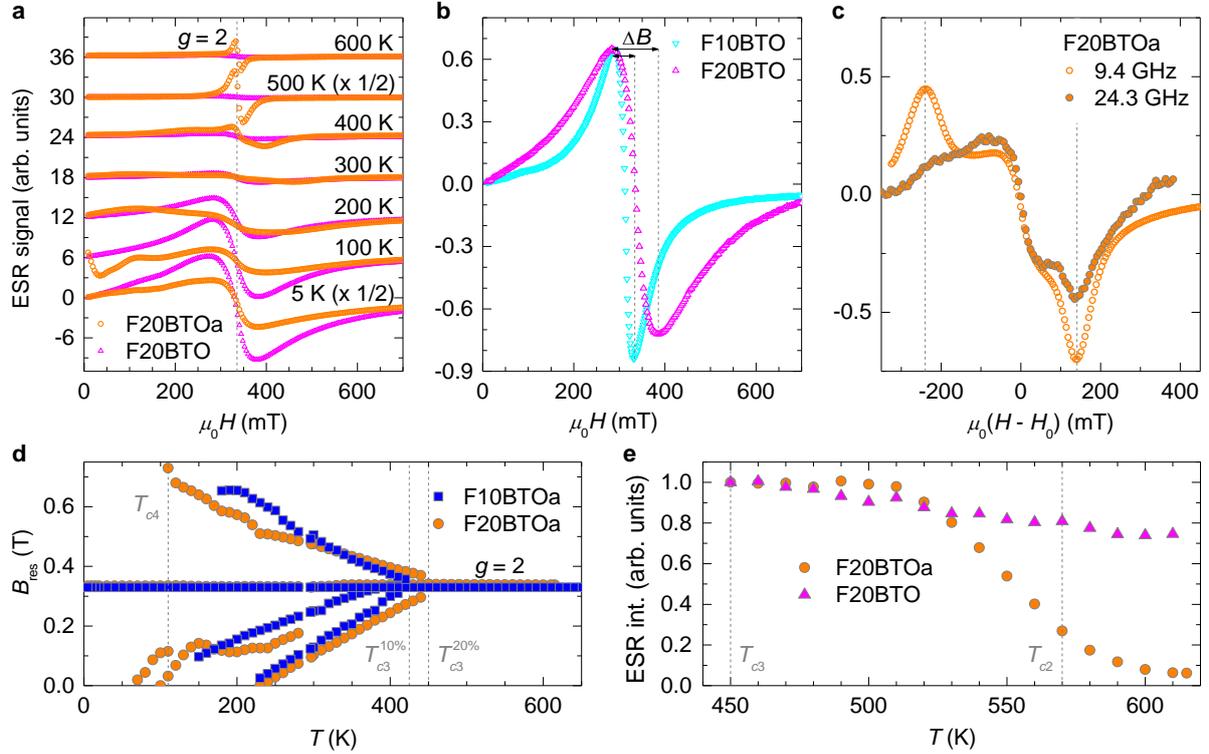

**Figure 3: ESR characterization of 6*H*-BaTiO$_{3-\delta}$.** (**a**) The temperature dependence of the ESR spectra in 20% Fe-doped 6*H*-BaTiO$_{3-\delta}$ displaying a single line in the non-annealed sample and a multi-feature line in the annealed sample. Spectra are displaced vertically for clarity. The $g$ = 2 position is marked with the vertical line. (**b**) The increase of the ESR line width with doping concentration in the non-annealed samples at 300 K. The peak-to-peak line width $\Delta B$ of both spectra is indicated by the horizontal arrows. (**c**) Comparison of the 300-K spectra of F20BTOa at 9.4 GHz and at 24.3 GHz. Side-band shifts that are frequency independent are marked with vertical lines. (**d**) Resonance field of various resonance modes observed in the annealed Fe-doped 6*H*-BaTiO$_{3-\delta}$ samples below $T_{c3}$. The temperature independent position (horizontal mode) corresponds to the $g$ = 2 mode. (**e**) Comparison of the ESR intensity decrease of the single-line

FM mode observed above $T_{c3}$ in F20BTOa and the PM signal of the F20BTO sample with Curie-like dependence. Both ESR intensities are normalized at 450 K.

In order to support the claim of sizable magnetic interactions we analyse the line width of the PM ESR spectra in the non-annealed samples. The peak-to-peak line width $\Delta B$ (Figure 3b) scales proportionally with the doping level ($\Delta B^{10\%} = 45(2)$ mT, $\Delta B^{20\%} = 91(2)$ mT). Such a linear increase is in accordance with the linearly increasing magnetic anisotropy and exchange interactions (see Methods), as deduced from the Weiss temperatures of both non-annealed samples. On the contrary, the dipolar interactions in the absence of isotropic exchange coupling[36] would yield much smaller line widths (see Methods). ESR thus verifies that the average exchange interaction $J$ among dopants is not negligible. Furthermore, the line width, being proportional to $c$, is a fingerprint of a uniform PM phase. Aggregation of magnetic ions into nano-scale clusters, on the other hand, would result in an inverse dependence of the ESR line width on $c$, as was recently observed[17] in Mn-doped $SrTiO_3$.

In sharp contrast to the simple ESR line shape of the PM non-annealed samples, the ESR spectra of the annealed samples are much more complex below $T_{c3}$. Here, they exhibit several well-displaced lines (Figure 3c) that shift significantly with temperature (Figure 3d). The field shift of the side bands from the $g = 2$ position is the same at 9.4 GHz and at 24.3 GHz (Figure 3c), which is indicative of static internal magnetic fields characteristic of the FM phase. Very importantly, above $T_{c3}$ a single-line spectrum is observed at $g \simeq 2$ (Figure 3a) in all the annealed samples despite the fact that the FM state persists to much higher temperature, $T_{c1}$ ~ 700 K. With its peak-to-peak line width $\Delta B = 9(1)$ mT, this spectrum is markedly narrower than the PM spectrum of the non-annealed samples. Moreover, its intensity strongly decreases between $T_{c3}$ and $T_{c2}$ (Figure 3e). The latter behaviour is in disagreement with the Curie-like decrease, characteristic of the PM non-annealed samples (Figure 3e), and is rather reminiscent of the FM magnetization decrease in the same temperature range (Figure 2a). Therefore, this $g \simeq 2$ mode is still a FM mode. We stress that its single-component line shape reveals that the local magnetic anisotropy, which dictates the shape of the ESR spectrum in the FM state[37] and depends on local symmetry, drastically changes at $T_{c3}$.

**Clustering of FM moments as evidenced by μSR**

The ESR measurements provide convincing evidence of FM moments in the annealed samples, as well as their absence in the non-annealed samples. In addition, they reveal a novel and important (vide infra) discovery about the local-symmetry reduction below $T_{c3}$. However, ESR does not yield any information about the spatial distribution of the FM moments. To answer this question we turn to the complementary μSR method (see Methods for introduction to this method). This local-probe technique is renowned in magnetic spectroscopy for its ability to detect even the smallest local magnetic fields at the muon stopping site and for the ease in which the experimentalist can discriminate between static and dynamical internal fields[38]. We find that in zero magnetic field (ZF), the time-dependent muon spin polarization $P(t)$ is monotonic at all temperatures in the experimental time window, independent of the PM/FM character of a particular sample (Figure 4a and 4b). It can be modelled with the stretched-exponential (SE) time dependence

$$P_{SE}(t) = e^{-(\lambda t)^{\beta}} , \qquad (1)$$

well suited for dynamical local fields in diluted spin systems[39]. Here $\lambda$ denotes the muon relaxation rate and $\beta$ the stretch exponent. For each sample the relaxation rate is found to increase notably with increasing temperature ($P(t)$ curves in Figure 4a and 4b decrease faster at higher temperature). This increase of the relaxation rate is very similar in the annealed and non-annealed samples for a given concentration of dopants. Therefore, the muon relaxation in the Fe-doped BaTiO$_{3-\delta}$ system is attributed to the fluctuating paramagnetic moments.

Direct confirmation of this statement is provided by comparing the ZF data and the data obtained in the weak longitudinal field (LF) of 1 mT. Such field comparison shows a very similar relaxation response (Figure 4c). We note that in addition to the dynamical model of equation (1), the Lorentzian Kubo-Toyabe (KT) model[38]

$$P_{KT}(T) = \frac{1}{3} + \frac{2}{3}(1 - \gamma_\mu \Delta_\mu t) e^{-\gamma_\mu \Delta_\mu t}, \qquad (2)$$

also provides a satisfactory fit to the ZF data. The KT model is applicable to disordered static local fields with the distribution width $\Delta_\mu = 7.0 \cdot 10^{-2}$ mT ($\gamma_\mu = 2\pi \cdot 135.5$ MHz/T is the muon gyromagnetic ratio). However, in this static case the 1 mT longitudinal field would almost completely remove the relaxation of $P(t)$, which clearly contradicts the experiment (Figure 4c). The muon relaxation due to local fields fluctuating with the exchange frequency $\omega_e = k_B J/\hbar \sim 1$ THz on the other hand, is not affected by such a small LF field[38], because $\gamma_\mu \cdot 1$ mT $= 0.85$ MHz $\ll \omega_e$.

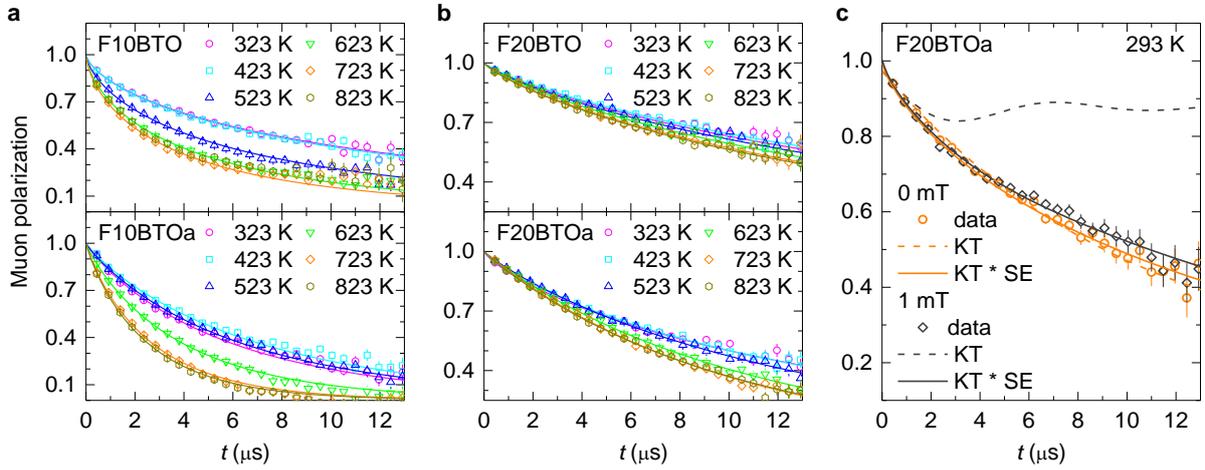

**Figure 4: μSR characterization of 6H-BaTiO$_{3-\delta}$.** The temperature dependence of muon polarization in (**a**) 10% and (**b**) 20% Fe-doped 6H-BaTiO$_{3-\delta}$ samples in zero applied field showing increased relaxation rates with increasing temperature. The upper panels correspond to non-annealed samples and the lower panels to annealed samples. The solid lines are fits to the stretched exponential (SE) model of equation (1). (**c**) The room-temperature muon polarization of the F20BTOa sample in ZF and in a weak longitudinal applied field. The dashed lines correspond to the static Kubo-Toyabe (KT) model of equation (2), while the solid lines are fits to the model encompassing both static KT and dynamical SE relaxation (see text for details). The error bars of muon polarization data are defined as a square root of the total number of detected positrons.

A closer look at the ZF and LF data of the F20BTOa sample at room temperature, however, reveals a somewhat decreased muon relaxation in the LF experiment (Figure 4c). This can be attributed to the presence of some weaker static relaxation in addition to the dominant dynamical one. A simultaneous fit of both datasets to the corresponding model $P(t) = P_{SE}(t) \cdot P_{KT}(t)$ yields the static-field distribution width of $\Delta_\mu = 6.1(3) \cdot 10^{-3}$ mT. This value represents the upper bound of the frozen-FM-moments' contribution in the annealed samples, since the static-field distribution may also partially arise from nuclear magnetic moments. However, for homogeneously distributed FM moments the calculated field-distribution widths at both oxygen sites, in the vicinity of which the muons are highly likely to reside in oxides[38], are almost an order of magnitude larger, $\Delta_\mu = 3.4(2) \cdot 10^{-2}$ mT (see Methods), even for the very small average value of the FM moments in F20BTOa ($1.5 \cdot 10^{-3} \mu_B$ at 300 K). As the interaction strength between the muon and the FM moment decreases with the cube of their distance, the much smaller experimental value of $\Delta_\mu$ can be reconciled by considering clustering of the FM moments. This thus provides microscopic evidence that, in contrast to the homogeneous distribution of PM moments evidenced by ESR, the FM moments are not uniformly distributed throughout the sample.

**Discussion**

The small size of the average FM moments and their non-uniform distribution in the annealed Fe-doped $6H$-BaTiO$_{3-\delta}$ samples cast doubt on the intrinsic origin of these moments. It is therefore crucial to critically assess the origin of ferromagnetism in this system. It could well be that the FM behaviour either corresponds to some kind of "impurity" phase or that the spinodal decomposition occurs and very small dispersed lattice-matched FM clusters of $Fe^{3+}$ ions form during the annealing process. The latter scenario has been recently demonstrated[17] in Mn-doped SrTiO$_3$. In order to check for the presence of such small clusters that would minimally deform the local crystal structure and would be practically undetectable even by high-resolution imaging techniques, we performed a magnetization relaxation experiment on the FM F20BTOa sample. After being exposed to the magnetic field of 1 T for 20 minutes, the subsequent remnant relaxation in zero field was measured. No magnetization relaxation towards zero has been detected on the time scale of several minutes (Figure 2c). This implies an unexpectedly slow relaxation process within the Stoner-Wohlfarth model of single-domain FM (nano)particles[40] and thus excludes the presence of small FM clusters of $Fe^{3+}$ ions.

Therefore, the alternative scenario of "impurity" ferromagnetism is much more plausible and should be carefully considered. The Curie temperatures of the most common iron oxides[41] that may develop as a by-product during the synthesis are much higher than $T_{c1}$ ~ 700 K; ~950 K ($Fe_2O_3$) and ~850 K ($Fe_3O_4$). On the other hand, the Curie temperature of ~680 K has been recently reported for the 3C bulk $BaTi_{0.95}Fe_{0.05}O_{3-\delta}$ sample, where a much larger magnetic moment of 0.75$\mu_B$ per Fe has been found[42]. The high-temperature ferromagnetism has regularly been encountered in this pseudocubic polymorph, even at much higher Fe-doping concentrations (up to 75%), where the 3C phase is stabilized only by the confined geometry of thin films and nanoparticles[43–46]. We note though that the Curie temperature of an impurity phase is usually size dependent for nano-particles, as well as it may depend on the level of doping. Therefore, the fair agreement of $T_{c1}$ with the Curie temperature of the 3C bulk $BaTi_{0.95}Fe_{0.05}O_{3-\delta}$ sample cannot be taken as a solid proof that the FM character of the Fe-doped 6H-$BaTiO_{3-\delta}$ is due to the impurity 3C phase. However, an unambiguous evidence that this is indeed the case comes from the slightly doping dependent $T_{c3}$ transition temperature ( $T_{c3}^{10\%} = 430$ K , $T_{c3}^{20\%} = 450$ K ; see Figure 3d), which exactly coincides with the doping dependent cubic-to-tetragonal structural transition of the 3C phase[43]. Moreover, this structural transition is clearly reflected in the FM ESR line-shape change from the high-temperature single line to a multi-feature spectrum below $T_{c3}$, which corroborates the change of the local symmetry from being highly symmetric above $T_{c3}$ to being less symmetric below $T_{c3}$. It is also worth noting that in epitaxially grown thin films, the 3C phase grows at the initial stages, while later on a disordered phase composed of 3C and 6H intergrowths forms[43]. Since increasing thickness decreases the FM response of thin-film samples[47], their ferromagnetism is most naturally attributed to the 3C phase, for which it is also theoretically predicted as the ground state[42].

All the above-presented arguments give clear evidence that the FM behaviour of the annealed Fe-doped $BaTiO_{3-\delta}$ samples is related to internal strains that locally destabilize the 6H phase in favour of the 3*C* phase, much as it happens globally in the confined geometries of thin films and nanoparticles. The μSR results yielding only minor differences between the FM annealed and PM non-annealed samples, as well as the non-uniform distribution of FM moments, corroborate this scenario. Furthermore, ferromagnetism arising from the sparse strain-stabilized defect regions with the 3*C* crystal structure explains both the extremely small magnitude of the average ordered Fe moment and its unusual decrease with increasing doping concentration; the increasing doping level being known to destabilize the 3*C* phase[35,48]. Finally, in the non-annealed samples that were heat-treated at lower temperature (see Methods for details) the crystallites are under significantly less strain than in the densely sintered annealed samples after additional annealing at higher temperature. Therefore, in the latter samples strong strain fields along crystallographically mismatched grain boundaries, or around plane defects, are likely regions for nucleation of stable pseudocubic domains with their own intrinsic ferromagnetism. These are not single-domain nano-sized FM defects, as magnetization relaxation is absent.

Our comprehensive magnetic investigation of various Fe-doped 6*H*-BaTiO$_{3-\delta}$ samples has thus revealed that the FM response observed in the annealed samples is not intrinsic to the hexagonal crystallographic phase, as broadly speculated before. This conclusion is based on our complementary bulk and local-probe magnetic investigations, providing new microscopic insight and extending the experimental temperature range far beyond all previous reports. Bulk magnetization measurements have demonstrated that extremely small average static magnetic moments develop in the annealed samples below $T_{c1} \sim 700$ K and several additional magnetic instabilities occur as the temperature is lowered further. Below the $T_{c3}$ transition (at around 430 K and 450 K in 10%- and 20%-doped samples, respectively) a significant symmetry reduction of the local structure is detected by ESR. The $T_{c1}$ and $T_{c3}$ transitions coincide with the ferromagnetic and ferroelectric transitions of the Fe-doped 3*C*-BaTiO$_3$ polymorph, respectively, while the microscopic characteristics determined by μSR imply non-uniformly distributed FM regions in the samples. This demonstrates that the FM response of the Fe-doped 6*H*-BaTiO$_{3-\delta}$ system originates from sparse regions where the pseudocubic structural polymorph is stabilized by strain fields. The strain-induced local competition between the two structural (and magnetic) phases is inherent only to densely sintered annealed samples. Such scenario of ferromagnetism may turn to be important for other DMO materials, where the competition between different structural phases is intrinsically present, in particular in confined geometries where strain effects are enhanced.

**Methods**

**Samples**

High quality 6$H$-BaTiO$_{3-\delta}$ polycrystalline samples with $c$ = 10, 20% Fe$^{3+}$ ions substituted for Ti$^{4+}$ were synthesized according to the procedure thoroughly explained in Ref. 28. For each composition, after heat treatment at 1250 °C, a part of the sample was additionally annealed in oxygen atmosphere at 1500 °C, typically for 5 – 10 hours. We label the non-annealed and annealed samples with F$c$BTO and F$c$BTOa, respectively. The variations of the annealing time showed no significant influence on the magnetic properties[28]. X-ray powder diffraction (XRD) was used to verify the single-phase hexagonal structure of all our samples. The elemental analysis was performed by Energy Dispersion X-Ray Spectroscopy (EDX) on polished ceramic surfaces with JSM-7100 F (Jeol) field-emission scanning electron microscope equipped with an x-ray detector (X-Max 80, Oxford Instrument). Ten EDX characterizations were performed on each sample and were statistically treated to obtain average values and standard deviations. The iron concentration in the nominally 10%-doped samples was found at 10.3±0.5%, while in 20%-doped samples it amounted to 19.9±0.6%. For all our samples the analytically determined compositions thus correspond to the nominal compositions within the small error bars.

**Bulk magnetization**

Bulk magnetization measurements were performed on a couple of Quantum Design SQUID magnetometers and a vibrating sample magnetometer (VSM) as a function of temperature (between 2 and 850 K) in various magnetic fields and at various fixed temperatures as a function of varying magnetic field. Zero-field-cooled (ZFC) and field-cooled (FC) temperature-dependent data were collected. A high-temperature insert was used for measurements above 400 K. The SQUID measurements were performed on samples sealed in quartz capillaries, while boronitride sample holders were used with VSM. The signal from the sample holders was carefully evaluated. The size of the ordered magnetic moment was estimated from the saturation value of the FM component in the hysteresis, by subtracting the linearly increasing PM part.

**ESR**

Electron spin resonance was measured with home-built resonator-cavity based X-band (9.4 GHz) and K-band (24.3 GHz) spectrometers. In the X band, a continuous-flow cryostat was used in the temperature range 5 – 300 K, while a preheated-nitrogen-flow heating system was used above room temperature up to 620 K. The samples were sealed in ESR silent quartz tubes. The ESR intensity was calibrated at room temperature with a reference sample ($CuSO_4 \cdot 5H_2O$). The ESR susceptibility of the non-annealed PM samples $\chi_{ESR}$ = 0.06(3) Am²/ T per mole Fe was found compatible with the bulk molar susceptibility $\chi_b$ = 0.086 Am²/T, proving that ESR was detecting the intrinsic signal of Fe dopants in $BaTiO_3$.

The second moment[36] of the ESR line $M_2$ arising from dipolar interactions is given by

$$M_2 = \frac{3}{4}\left(\frac{\mu_0}{4\pi}\right)^2 S(S+1)(g\mu_B)^4 \sum_k \frac{\left[3\cos^2\theta_{jk} - 1\right]^2}{r_{jk}^6} \quad (3)$$

where $\mu_0$ is the vacuum permeability. The sum runs over all neighbours $k$ of a given site $j$, connected by the vector $\mathbf{r}_{jk}$. $\theta_{jk}$ is the angle between $\mathbf{r}_{jk}$ and the applied magnetic field. $M_2$ was calculated by powder-averaging equation (3) for a lattice fully occupied with $Fe^{3+}$ ions on either the Ti(1) or Ti(2) site. In the limit of negligible exchange interactions $M_2$ yields the ESR peak-to-peak line width $\Delta B \simeq \sqrt{M_2}/(\sqrt{3}g\mu_B)$ for the fully occupied lattice and the line width $\Delta B(c) = c \cdot \Delta B$ in the dilute limit[17,36]. We find $\Delta B_{Ti(1)}^{20\%} = 23$ mT and $\Delta B_{Ti(2)}^{20\%} = 31$ mT, which is far below the experimental values. In the exchange narrowing limit ($k_B J \gg g\mu_B \Delta B$), on the other hand, the ESR line width is given by $\Delta B \simeq M_2/g\mu_B k_B J$. The second moment $M_2$ is quadratic in the anisotropic interaction (i.e., in the doping level $c$), therefore $\Delta B$ also scales linearly with $c$ for a linearly dependent $J$.

**μSR**

In a μSR experiment almost 100% spin-polarized muons are implanted into a sample. As particles possessing a magnetic moment, they interact with local magnetic fields, which drive the time dependence of muon polarization $P(t)$. At the time of the muon decay a positron is emitted preferentially in the direction of the muon's spin direction. The detection of positrons therefore allows for the reconstruction of $P(t)$ and thus for the determination of a magnitude, distribution, and fluctuation rate of local internal magnetic fields.

Muon spin relaxation measurement were performed on the EMU instrument at the ISIS facility, Rutherford Appleton Laboratory, UK. A flat-plate furnace was used to cover the temperature range between 290 and 820 K. The background signal from a titanium sample holder was estimated with a hematite reference sample. It was found to typically amount to ~30% of the total signal and was subtracted from the μSR data shown above. The measurements were carried out with detectors grouped in the forward-backward direction with respect to the muon beam in zero field (ZF) and in longitudinal applied magnetic field (LF), while calibration measurements were performed in a weak transverse field of 2 mT.

The dipolar-magnetic-field distribution at a given muon stopping site, $D(B) = \frac{1}{N_0}\frac{dN}{dB}$, can be accurately modelled, as demonstrated in Ref. 49. In $6H$-BaTiO$_{3-\delta}$ it was calculated by randomly populating fraction $c$ of all Ti sites with frozen Fe$^{3+}$ moments. The magnetic field at a chosen crystallographic site was then calculated in $N_0 = 2,048$ different crystallographic cells. For each point all magnetic moments within a spherical region around it large enough to ensure convergence were taken into account.

## Acknowledgements

We acknowledge the financial support of the Slovenian Research Agency Programs P1-0125, P2-0377, P1-0112 and Projects N2-0005, Z1-5443, as well as the Research Support Programme of University of Zagreb. The μSR study has been supported by the European Commission under the 7th Framework Programme through the 'Research Infrastructures' action of the 'Capacities' Programme, Contract No. 283883-NMI3-II.

## Author contributions

A.Z. and M.V. designed and supervised the project. I.A., I.M. and M.V. synthesized and structurally characterized the samples. Z.J. and D.P. performed bulk magnetization measurements. M.P. and M.G. conducted ESR measurements. A.Z., M.G. and M.T. carried out the μSR study. All the experimental data were analysed by A.Z., who also wrote the paper. All authors discussed the results and reviewed the manuscript.

## Additional information

### Competing financial interests

The authors declare no competing financial interests.